# *Stardust* findings favor not only the planetary origin of comets but the underlying close-binary cosmogony of the Solar system as well

Edward M. Drobyshevski

*Ioffe Physical-Technical Institute, Russian Academy of Sciences, 194021 St.Petersburg,Russia*
*E-mail address:* emdrob@mail.ioffe.ru

**Abstract**

The paper analyzes findings of the *Stardust* mission that brought to the Earth dust from the 81P/Wild 2 coma. Just as the data obtained in the *Deep Impact* mission to 9P/Tempel 1, they are at odds with the universally accepted condensation/sublimation cometary paradigm. They fit rather well to the approach assuming ejection of nuclei of short-period comets from moon-like bodies of the type of Galilean satellites in rare (six to seven events in 4.5 aeons) global explosions of their massive icy envelopes saturated by $2H_2 + O_2$, products of the electrolysis of ice. This approach offers an explanation, in particular, for the jet activity of comets, which is sustained by combustion of the $2H_2 + O_2$ + organics mixture ignited and complemented by the solar radiation. Combustion accounts also for other observations, in particular, the presence in the dust of products of high-temperature (800-900 K) metamorphism. The presence of minerals forming at still higher temperatures (~1400-2000 K), just as the undoubtedly planetary origin of some long-period comets arriving from the *joint planeto-cometary cloud* beyond Neptune, forces one, however, to invoke the close-binary cosmogony of the Solar system, which three decades ago predicted the existence of such a cloud (in the recent decade, this prediction has been substantiated by the discovery there of many dwarf planets). This cosmogony is based on the modern understanding of the processes involved in the formation of multiple stellar systems and of their gas-dynamic evolution. It considers the Jupiter-Sun system as the limiting case of a binary star and uses it as a basis for explanation of all the known observations and for prediction of the new ones to come. It provides a plausible explanation, in particular, for both the origin and capture by the Earth of the Moon as a high-temperature condensate and the formation of the Galilean satellites, which also contain inclusions of refractory minerals in the ices of their envelopes.

# 1. Introduction

*Deep Impact* and *Stardust*, active experiments aimed at exploring properties of cometary nuclei and of the material escaping from them, have challenged the traditional condensation-sublimation paradigm treating comets as a dead product of accretion of matter that had condensed in the outer cold fringes of the Solar system. Actually, this paradigm had earlier been likewise incapable of explaining the origin of outbursts and fragmentation of cometary nuclei, jet outflow of matter from the nucleus and the appearance of radicals and ions close to the nucleus etc., without violation of the conservation laws and invoking additional hypotheses. This paradigm has rather been resting, in accordance with the concepts of Kuhn (1962), on the inertia of thought, an essential mechanism that keeps sustaining the living organism of science. It has suddenly turned out (e.g. Weaver, 2004) that a cometary nucleus is not a friable lump of snow or loose rubble pile of ~100-200 m size dirty ice 'grains', which are ready to break up under the tidal action of the Jupiter into thousands re-assembling pieces as it was assumed by Asphaug and Benz (1994) in their frequently referred paper (non-tidal scenario of the SL-9 post-perijove splitting is given by Drobyshevski, 1997). The nucleus surfaces demonstrate regions of different morphology and geological processing, layering, numerous large-scale structures, including ones resembling impact craters, etc. (A'Hearn et al., 2005; Brownlee et al., 2004; see also Sec. 4).[1] Such rather vast coherent formations cannot be created in absence of significant gravity or strength of material binding all parts of nucleus together. The comet ices contain rocky particles that had been acted upon by high temperatures. The presence of particulate products of high-temperature condensation and metamorphism has caused a certain consternation and bewilderment which ended for the moment in invoking new and contradictory hypotheses assuming strong mixing in the preplanetary disk (indeed, turbulent mixing shortens sharply the disk lifetime, an argument that, when applied to standard approaches, creates serious difficulties for the formation of giant planets) and/or ballistic ejection of particles out of the intra-mercurian zone into Neptune's zone and beyond it (which would require velocities above 60 km/s!).

Strange as this might seem, the latter point follows directly from our concepts of the Jupiter-Sun system as the limiting case of a binary star, and of other planets as a by-product, albeit unavoidable, of its early evolution (Drobyshevski 1974b, 1978, 1996). This is what can be called the close binary cosmogony (CBC) of the Solar System. It does not contain any additional hypotheses and may be considered as a natural consequence of the scenario of binary star formation developed by us, which rests on sufficient statistics and modern ideas of the gas-dynamic evolution of binary stars (Drobyshevski 1974a, 1976). A brief exposition of the fundamentals of CBC and of a number of its implications which have been enjoying in the recent decades substantiation by observations is presented in Sec. 2. Section 3 considers also the inferences of CBC bearing on the planetary origin of the short- and long-period (SP and LP) comets. These comets are an inevitable endpoint of evolution of various groups in the great number of small (Moon- and Pluto-like) planets that had formed, just as large planets, in the proto-Jupiter (PJ). SP comets are produced in global explosions of icy mantles of Galilean satellites and of Titan that are saturated by products of ice electrolysis (the New Eruptive (Explosive) Cosmogony of comets, NEC), and LP comets, in rare collisions of trans-Neptunian dwarf planets and, further on, of their fragments. Section 3 reminds us also of the fact that it was the NEC that offered a rational explanation for many of the unexpected findings of *Deep Impact* mission. Section 4 analyzes the observations amassed in the *Stardust* flyby past the nucleus of 81P/Wild 2. These are the images of the nucleus and data on the frequency of impacts, the mass and composition of the incident particles. Many of these results turned out a surprise for the specialists. They were, however, expected and are interpreted in a straightforward way in terms of the NEC. Section 5 stresses that the material retrieved by *Stardust* exhibits invariably indications of having been subjected to high temperatures, in some cases even in the free oxygen presence, which implies combustion in the jet sources that was predicted by the NEC. It was also pointed out that it is the compact (primarily mineral) grains that belong to the primary cometary material, whereas the fluffy particles, in contrast to traditional ideas, form in the combustion products of electrolyzed ices. The issue of the origin of the Ca-Al-rich inclusions and of other high-temperature metamorphized minerals in the nucleus of Wild 2 is addressed in Sec. 6. The temperature $T \approx$ 1400-2000 K required for their appearance defies explanation within traditional approaches. If we accept, however, that comets are actually ejecta from planets that had formed, as follows from the CBC, in the convective (turbulized) PJ, within which the temperature varied from $\sim 10^2$ to $\sim 10^4$ K, this problem finds a ready solution. Finally, in Sec. 7 we conclude by pointing out that the hopes of the scientists who initiated the cometary missions that the latter would shed new light on the origin of the Solar System have been fully justified. Missions to the LP comets and, particularly, to Callisto to measure the concentration of $2H_2 + O_2$ in its ices appear to be highly promising and rewarding.

---

[1] It is instructive here to compare these real images with the pre-encounter drawing by Pat Rawling of a comet nucleus as a rather loose conglomerate of the ~100-200 m size 'grains' (see the front cover in Russell, 2005).

## 2. Close-binary cosmogony of the Solar System and trans-Neptunian planeto-cometary cloud

The CBC is a viable alternative to the traditional disk paradigm of the origin of the Solar System and of the planets and comets it contains. The latter paradigm is not capable of accounting for many old and, all the more so, new facts and is burdened by a large number of *ad hoc* hypotheses. The CBC starts with the idea that planetary systems themselves, including the Jupiter-Sun system, are actually the limiting case and/or by-product of the processes involved in the formation and early evolution of binary stars. Such concepts could not have been put forward two centuries ago, in the times of Kant and Laplace, founders of the disk cosmogony, and even half a century ago, when Gamow (1945) was feeling that the old cosmogony is unphysical but could not offer a viable alternative in its place, because many essential aspects of the evolution of stars and, much less so, of their multiple systems were far from clear (Drobyshevski, 1996).

Close binaries form in the fission of the last proto-stellar fragment, the product of multi-stage fragmentation of a rotating collapsing molecular cloud. It breaks up into components as a result of the onset of rotational-exchange instability (Drobyshevski 1974a, 1976, 1996, 1999).

This rapidly collapsing fragment forms close to the limit of rotational stability, with its inner regions, because of their larger density, rotating faster than the outer regions do. When Hayashi convection sets in such a protostar, the convection mixing transports the angular momentum rapidly outward. A massive ring separates from the protostar. The ring is unstable against azimuthal fragmentation and breaks up into several self-gravitating bodies. They interact with one another to coalesce eventually or become ejected out of the system (the problem of more than three bodies; see Fig. 1 in Drobyshevski, 1996). The first Lagrangian point for the remaining, most massive (or lucky) fragment (probably moving in an elongated orbit, which could account for the large orbital eccentricities of many exo-Jupiters discovered in the recent years) lies within this protostar with persistent convection (the gas polytrope index is $\approx 3/2$). Therefore, all of its convecting material flows on the dynamic time scale over onto the newly formed object. If the protostar mass is $\geq 3 M_\odot$, convection persists only within an outer part of its body, and subsequent exchange of material ends up in formation of a pair of components with close masses, a scenario that explains well enough the statistics of close binaries among the F, A, and B stars (Drobyshevski, 1976). If, however, the starting mass is ≤$1.5M_\odot$, convection extends over all of the protostar, so that one would expect all of its mass to transfer to the newly forming object. The non-adiabaticity of the gas expansion inside the component that is loosing gas and the onset of water condensation in its gas put an end to the overflow. By this time, only about ~$10^{-3}$ of the starting mass of the parent protostar is left: this is how the Jupiter was borne. And the lucky fragment of the broken up ring to which the parent protostar has transferred almost all of its mass, becomes the Sun (Drobyshevski, 1974b).

Saturn could be the remainder of a not so lucky fragment of the broken up ring, while the terrestrial planets and the rocky nuclei of Uranus and Neptune appeared in the massive (≤$0.1M_\odot$) PJ which overfilled its Roche lobe, with the material flowed through the first Lagrangian point onto the proto-Sun. The rapidly rotating PJ is an extremely dense analog of the classical (preplanetary) disk. The very fast agglomeration of the condensing material into planetesimals and, further, into planets occurred here at densities and temperatures (surface density ~$10^5$ g/cm$^2$, $T \leq 5000$ K) inconceivable for conventional preplanetary disks. Numerous bodies thus formed moved within the PJ at relatively low velocities (≤1 km/s) and suffered high gas resistance, the conditions that differed favorably from those typical of traditional heliocentric disks and did not interfere with coalescence of the planetesimals and easy capture of satellites. This scenario easily solves, in particular, the problem of the appearance of the Earth-Moon system (where the Moon could be from the outset a product of a high-temperature condensation in the PJ hot regions rather than of a giant impact, while the latter possibility could not be excluded in these conditions; Drobyshevski, 1975) and possibly of the Neptune-Triton system. Because of the fast (it took about $10^3$ years) reduction of the PJ mass, these bodies moved in it along unwinding trajectories and, thus, left eventually the PJ. The latter bodies that had formed already in the adiabatically cooled PJ could not leave it because it sank into its Roche lobe and stopped loosing mass. These are Galilean satellites with the originally high (~50%) ice content. Their dirty ices consist of water and "primitive" organics, as well as contain numerous mineral inclusions in the form of dust and rocks, the products of diverse thermal metamorphism (this relates to the organics as well), agglomeration, and mutual collisional comminution in the convective gaseous body of the PJ (Drobyshevski, 1978).

Galilean satellites offer us a characteristic lower mass scale for bodies capable of escaping from the PJ - these are bodies of about lunar mass. It is becoming increasingly more obvious that they define apparently a certain maximum limit on the lowest mass scale. Indeed, in late stages the rate of the PJ mass decrease began to slow down, so that it is this period of the evolution that can be identified with the escape of bodies of lunar (and larger) mass out of the PJ. Earlier mass loss rates should have been two to three times higher, and, therefore, the bodies that could leave the PJ at that time should (for the same average material density) be smaller in size than the Galilean satellites by a factor of two to three (see Drobyshevski, 1978). All in all, ~$10^4$ moonlike bodies (MLBs) (Drobyshevski, 1978) or ~$10^5$ bodies ~$10^3$ km in size could have formed in the PJ and left it together with their satellites. Because the above estimates are valid to within an order of magnitude, we are going to refer in what follows to all such objects as MLBs.

Most of these ~$10^4$-$10^5$ MLBs were swept out of the Solar system by mutual perturbations and those of the large planets. Uranus and Neptune with their atmospheres are apparently also members of the planetary ensemble that were borne in the PJ and, later, interacted with this MLB "gas" to drift away from the Jupiter (and Saturn) to their present position (Drobyshevski, 1978; see also, for instance, Fernandez and Ip, 1984; Hahn and Malhotra, 1999, 2005; Gomes et al., 2004, who consider the drift of giant planets in a massive disk of planetesimals which do not interact with one another). In analyzing the evolution of this viscous "gas" consisting of the Coulomb interacting MLBs (coined presently by "dynamic friction"), one has to keep in mind that in the initial stage their motion was perturbed not by the compact, practically point-like Jupiter but by a gaseous PJ, which initially was spread out over its Roche lobe ~1-1/4 AU in size and subsequently, after becoming confined under its surface, contracted gradually because of cooling. Therefore, the resultant perturbations by the PJ were originally relatively weak; also, the MLBs themselves could cross it losing on the way a part of their kinetic energy. The smallest bodies slowed down by the gas could re-evaporate or settle down to the PJ center. The resistance of the gas dissipating into space from the flow of matter streaming out from PJ to the proto-Sun could also play some role in the initial evolution of the MLB ensemble. None of these factors should be overlooked in studying the evolution of an ensemble of interacting MLBs by numerical modeling in the way this was done by Chambers (2001), Morbidelli et al. (2002), Fetbroit (2002), etc.

A fraction of the mutually interacting MLBs ($\sim 10^3$-$10^4$, depending on the characteristic mass scale of the bodies that had escaped from the PJ) retained and crowded in the not too far periphery of the Solar System (~50-300 AU) to form there a quasi-toroidal cloud of dwarf planets moving in disordered orbits (Drobyshevski, 1978). It appears only natural to class Pluto among them. Having analyzed the Uranus and Neptune axial tilts and the conditions that governed the capture of Triton, as well as of Charon, Stern (1991) also came, shall we say, to an "observational" conclusion drawing on verifiable material which concerns the past presence of numerous ($\sim 10^3$-$10^4$) MLBs in the 20-50 AU region. This consequence of the CBC has found substantiation in the recent decade in the discovery of objects of the type of Sedna, Eris, etc. Classical disk cosmogonies have revealed their inability to make such a prediction, which accounts for the present consideration of the MLB dynamic evolution being performed in *a posteriori* fashion (see, for instance, Duncan and Levison, 1997). Moreover, in an attempts at considering the origin of bodies within the Edgeworth-Kuiper belt (EKB) in terms of *in situ* formation, the authors of such papers have sometimes to invoke hardly justifiable assumptions of originally extremely small orbital inclinations and eccentricities of the coalescing blocks (see, e.g., an interesting paper by Morbidelli et al., 2002 and refs. therein).

## 3. On the origin of comets

*3.1. Long-period comets and trans-Neptunian planets*

The presence of a multitude of planetary bodies on distant disordered orbits sheds light on the scenario of formation of the LP comets. These are icy fragments produced in extremely rare direct collisions among MLBs, which entail partial disruption of the latter. This is how the present-day *joint planeto-cometary cloud* had formed (Drobyshevski, 1977, 1978).

This is neither the quasi-spherical Oort cloud extending to distances of up to $R \sim 10^5$ AU, nor the EKB ($R \sim 40$-$50$ AU), nor again the scattered comet disk postulated recently (see, e.g., Torbett, 1989, and the reviews by Morbidelli and Brown, 2004 and Duncan, Levison and Dones, 2004). All these formations, by their initial definition, should not contain planet-like bodies large enough that their gravitational perturbations could affect the orbital evolution of small bodies (including cometary nuclei) and drive them into the region of influence of Neptune and of other known planets. It is believed that objects populating the Oort cloud are perturbed by stars and/or Galactic tides, while motion of the EKB or scattered disk objects is influenced by Neptune's perturbations. A question immediately arises, for instance, of what could drive a Pluto-like Sedna into an orbit with a perihelion of 76 AU or Eris with its orbit inclination of 44º, short of a close encounter with another body, a member of the same planetary cloud. (One might cite here an interesting recent paper by Gomes, Matese and Lissauer (2005) who considered the effect of even not too large a planet with $\sim M_\oplus$; one might invoke also, as this was usually the case with the Oort cloud, stellar perturbations (see, e.g., Morbidelli and Levison 2004).)

And only the discovery in these outer reaches, considered traditionally to be exclusively a cometary domain, of planetary bodies forces researchers to accept *de facto* the existence of such objects and to study their behavior (see, e.g., Duncan and Levison, 1997); a hard task indeed would it be to explain their formation here and evolution within the framework of classical nebular concepts (see, e.g., Morbidelli et al., 2002; Brunini and Melita, 2002; Morbidelli and Brown, 2004 and refs. therein). For example, the observed deficiency of bodies in the scattered EKB disk combined with a lack of order in their orbits represents a certain problem (see, e.g., Gomes, 2003; Hahn and Malhotra, 2005). For people adhering to the concepts of disk cosmogony, which assumes *in situ* formation of bodies, these should be mutually exclusive and contradictory findings. This contradiction is, however, removed if the bodies are assumed to have got there, including mutual perturbations (it is worth to note that Hahn and Malhotra, 2005 also came to similar conclusion), from the PJ region, i.e., from within Neptune's and Uranus' orbits, that also drifted outward due to interaction with them, as this was conjectured after our 1978 publication by Fernandez and Ip (1984), who recognized and attempted to overcome the difficulties entailed by the *in situ* formation of Uranus and Neptune in the disk. Their results are in line with our inferences concerning evolution of planetary ensemble begotten by PJ. It appears appropriate to repeat here once more that the conclusions drawn in 1977 and 1978 from the CBC have been enjoying substantiation in the recent decades. (It is instructive to note also that the mass of the planetesimal disks invoked presently to account for the outward drift of Uranus and Neptune and creation of EKB is about $\sim 10\text{-}100 M_\oplus$ (see, e.g., Hahn and Malhotra, 1999), which is in accord with the total mass of the MLBs that by the CBC could have formed and escaped out of the PJ.)

Therefore all the ensembles (initially assumed cometary) like the EKB, Oort cloud or the scattered disk are considered within the CBC as somewhat spaced components of a *joint planeto-cometary cloud* that formed in the early stages of the close-binary Jupiter-Sun system. Its evolution was (and partially continues to be) governed by planetary bodies of the type of Pluto and, judging from the orbital parameters of Sedna or Eris, even by larger ones, just as this had been assumed from the very beginning by Drobyshevski (1978). Neptune, as one of many bodies formed in the PJ, is in no way an exclusion, but, as a result of having captured and retained a large gaseous atmosphere, its mass is large, and it separated out of this planeto-cometary cloud a group of objects, which are strongly bound to it and include, besides comet- and asteroid-like bodies, Pluto-like planets as well.

It is hard to understand why, despite a decade-long analysis by a number of researchers (e.g., Stern, 1996; Kenyon and Luu, 1999) of the part played by collisional (erosion) evolution, say, of the most massive and gravitating objects (i.e. Pluto-like dwarf planets) in the formation of the EKB and accounts of the existence there of small fragments due also to these collisions, seemingly nobody has thus far pointed out that as such ice fragments should behave as LP cometary nuclei, so that at least some of the LP comets should have practically present-day *planetary origin*!

True, a decade ago Farinella and Davis (1966) studied a possibility of appearance of SP comet nuclei in cascade collisions of asteroid-sized EKB objects and, further on, of their fragments, with resulting collisional ejection into dynamical resonance with Neptune. On the other hand, Chiang (2002) and Brown et al. (2007) revealed an existence in EKB of collisional families of (icy) objects created by total or partial disruption of dwarf planets. Of course, to be collisionally or gravitationally ejected into high-inclined orbits, the LP nuclei have to be mainly a result of collisions of planets orbiting well beyond the classic EKB (at ~$10^2$ AU or more), i.e., in the main region of the planeto-cometary cloud.

*3.2. Short-period comets and the NEC*

Another consequence of the CBC is the large original mass content of ices (~50%) on all Galilean satellites. Observations do not reveal ices on Io, their content on Europa is <10%, on Ganymede - ~43%, and on Callisto, ~50%. This difference cannot be accounted for within the frame of conventional ideas of their condensation in the disk that surrounded the early hot Jupiter (see Drobyshevski, 1980b and refs. therein).

A simple and straightforward explanation for these and other features of Galilean satellites as well as for a multitude of other objects and phenomena (see the list and refs. in Drobyshevski, 2000 and Drobyshevski, Kumzerova and Schmidt, 2006), including the SP comets, is proffered by the NEC of comets that considers the consequences of the inevitable large-scale volumetric electrolysis of ices with embedded mineral and carbonaceous (organic-like) inclusions in the massive icy envelopes of distant Moon-like bodies (Drobyshevski, 1980a). The electric current needed for the electrolysis was generated as, for instance, the Galilean satellites passed in their orbiting through the strong primeval magnetic field of Jupiter (Drobyshevski 1980b). The products of the electrolysis, $2H_2 + O_2$ mainly, build up in the ice in the form of a solid solution (~clathrate). When their concentration reaches ~15 wt%, such a solid solution becomes prone to detonation (see Drobyshevski et al., 2005, 2007, and refs. therein). Global explosions of the icy envelopes culminated in the loss of ices on Io (two to three explosions), Europa (two explosions), and, partially, on Ganymede (one explosion). Callisto's ices have not exploded thus far. The ices on Titan seem to have exploded only ~$10^4$ years ago (the dating was done based on an analysis of the specific features in the orbits of young LP comets of Saturn's family (Drobyshevski, 1981, 2000)), an event that created its atmosphere, Saturn's rings etc. and has prompted a number of predictions many of which have already found confirmation (for more details on this, see Drobyshevski, 2000). The kilometer-sized chunks of the outer layers ejected in the explosions become nuclei of SP comets which replenish their reservoirs in the giant planets' zones (with a part of them entering the LP orbits, which are naturally crossing orbits of the parent giant planets, Jupiter or Saturn). Their ices are also saturated by the products of electrolysis (a feature distinguishing them from the true (original) LP cometary members of the *joint planeto-cometary cloud*, which are borne in collisions among its planetary members). They are capable of sustaining exothermal reactions (burning, detonation) if an additional energy (insolation, impact) is imparted. This accounts for many manifestations of the SP comets, such as the jet character of dust-carrying gas (products of combustion) emerging from small-area sources on the surface of the nucleus, the appearance in jets in the immediate vicinity of the nucleus, where photolysis has not yet set in, of radicals and ions, an excess of CO over the $CO_2$ content, and of submicron-sized CHON particles, a feature characteristic of the burning of organics under oxydizer deficiency etc. (Drobyshevski, 1980a, 1988 and refs. therein).

In view of the well-known difficulties involved in the transfer in low-inclination orbits of observed number of SP comets of the Jupiter family from trans-Neptunian region (e.g., Duncan et al., 2004; Farinella and Davis, 1996; Torbett, 1989), one cannot rule out the possibility that it is because of a recent explosion on Titan that we live presently in the epoch of an excessive flux of SP comets (ideas of the present number of SP comets being excessive were voiced earlier; see, for instance, Napier and Clube, 1979).

The differences in the origin and, hence, in the composition of the LP (collisions of bodies in the planeto-cometary cloud) and SP (fragments of exploded electrolyzed ices) comets account for those in their manifestations; indeed, the activity of the former reaches usually a maximum before perihelion (the surface ices have already sublimated, as a rule, at reaching perihelion), and that of the latter, after the perihelion, when burning in local nests with an excess $2H_2 + O_2$ concentration becomes increasingly more intense as the ices warm up (one has naturally to recall that planetary perturbations are capable of transporting some LP comets in the orbits of the SP comets, and vice versa).

*3.3. DEEP IMPACT and the NEC*

NEC concepts provided a firm basis for our predictions and explanation of many results gained in the *Deep Impact* mission to the 9P/Tempel 1 comet (Drobyshevski et al., 2005, 2007). In particular, we have succeeded in "saving", in a certain sense, the energy conservation law. Indeed, an impactor 372 kg in mass with a velocity of 10.2 km/s carried an energy $E_0 = 19.6$ GJ. The mass ejected from the impact crater was, however, $M_{esc} \geq 10^7$ kg and it flew out with a velocity, at the lowest, of a few hundred m/s (in real fact, ~0.2-1 km/s). (Incidentally, in contrast to the first estimates of $M_{esc}$ as $10^6$ kg made by the people engaged in the experiment, we showed that the real value is at least an order of magnitude larger. In a recent publication, A'Hearn (2006, December) also gives $M_{esc} \sim 10^7$ kg.) The impactor struck apparently the ice that contained, besides the organics, $2H_2 + O_2$ as well, so that it initiated in this mix forced detonation with a liberation of energy exceeding $E_0$ by a few times at least. Forced detonation sets in at its threshold, i.e., at $T \approx 900$ K. It turned out that it is this temperature that matches the observed leading edge velocity of the water vapor outflow into vacuum, as well as the onset of transformation of amorphous to crystalline silicates, which were detected in the impact-produced plume (for more details, see Drobyshevski et al., 2007). The density of the Tempel 1 nucleus estimated from the *ballistic flight time* of the ejecta brought in to surface was found to be somewhat larger than expected (620+470/-330 kg/m$^3$; A'Hearn et al., 2005). It is worth of noting that these estimations are strongly model-dependent, so that researchers involved in the *Deep Impact* and *Stardust* projects are traditionally inclined to prefer lower values of density (e.g., ~400±300 kg/m3; Richardson and Melosh, 2006). However, if the ejection was indeed extended in time by an additional energy release, the density may rise accordingly to and even exceed 1000 kg/m$^3$, so matching the density of fragments of dirty monolithic ice with mineral inclusions. Which implies that these results would be in agreement with the NEC in this case too.

**4. The findings of *STARDUST* and their implications**

Mission *Stardust* made a rendezvous with comet 81P/Wild 2 at a distance $R = 1.86$ AU from the Sun, when the spacecraft flew in January 2004 past its nucleus at a distance of 236 km with a relatively low encounter velocity of 6.1 km/s (A'Hearn, 2006).

The experiment consisted actually of two phases, namely (1) obtaining 72 images of the nucleus and sampling of the cometary dust flux intercepted by the detectors in particle size and composition distribution, and (2) making a comprehensive laboratory analysis of the composition of the particles that were captured by a silicon aerogel trap 1048 cm$^2$ in area or left impact craters in a ~100-cm$^2$ Al foil strips and brought to Earth in January 2006 (Brownlee et al., 2006).

The images (Brownlee et al., 2004) revealed that the nucleus of Wild 2 is aspherical (an oblate ellipsoid with the radii of 1.65×2.00×2.75 ± 0.05 km) with distinct manifestations of a layered structure. The surface has depressions, part of which may be identified as impact craters, while others are of a different nature. The walls of the latter, up to 140 m high, are in some cases very steep (>70$^o$) and even with overhanging rims. One observed even isolated pinnacles with a height of up to 100 m. All this is at odds with the standard notions of the nucleus as a fluffy formation and suggests a noticeable strength of the material (Weaver, 2004).

One detected about 20 strongly collimated jets, possibly bursting forth from small subsurface sources at different angles to the surface. (We are witnessing here the old problem, dating back to the times of P/Halley (Drobyshevski, 1988), of the difficulties involved in concentrating the required amounts of solar radiation into small and essentially subsurface regions propelling gas-dust jets to a velocity of ~1 km/s, which is considerably in excess of the probable velocity of sublimating products of ~0.5 km/s, i.e., the problem of the source of energy to sustain jet-like sublimation of ices.) Two jets were found to be streaming on the shadowed side of the nucleus, and 16 of them, from local slopes.

Sekanina et al. (2004) pointed out that the gas jet expanding sideways causes separation of the carried dust particles in mass, because small particles are easier to be dragged by the gas. The data amassed by the Dust Flux Monitor Instrument (Tuzzolino et al., 2004) revealing short transits of the spacecraft through quite frequently anti-correlated regions with particles of different masses ($m \sim 9.8\times10^{-12}$ g and $m > 3\times10^{-8}$ g) provide grounds for such an inference. Besides, the data collected argue apparently in favor of slightly different size distributions of small and large particles (with a relative excess of large particles?), which may imply their being of different origins. The approach developed by Sekanina et al. (2004) would suggest that as the spacecraft crosses a jet, one should first detect at its boundary impacts of small particles, at the center, i.e., near the jet axis, larger particles, and at the opposite outer boundary, small grains again.

Experiment shows, however, that this is not quite so. For instance, entering the jets $\rho$, $\tau$, $\mu$, and $\upsilon$ was indeed accompanied by recording a large amount of fine dust (at the detector sensitivity limit, primarily $m \leq 10^{-10}$ g), whereas at the exit from the jet the instrument detected only particles with $m \geq 10^{-8}$ g (see Fig. 2 in Sekanina et al., 2004). The appearance of small particles is usually attributed to (photolytic?) fragmentation of larger-grained dust. We would suggest, however, the opposite interpretation, which provides also an incidental explanation for the absence of fine particles at the trailing edges of these jets. It suggests clumping of invisible (and undetected) particles of amorphous carbon and organics, which are products of incomplete combustion, ~0.01-0.2 µm in size, in detectable (>1 µm) flakes of soot and smoke, as this usually occurs in incomplete combustion (e.g., Palmer and Cullis, 1965).

The matter is that the probability for small particles to coalesce is proportional to the square of their volume concentration, i.e., it grows steeply when expanding boundary zones of neighboring jets come in contact, interpenetrate and become compressed by arising shock waves. Figure 2 in Sekanina et al. (2004) provides supportive evidence for this suggestion; indeed, the maximum fluxes of fine dust are seen to favor areas of possible contact of the side boundaries of the abovementioned $\rho$, $\tau$, $\mu$, and $\upsilon$ jets, as well as those where the $\alpha$, $\beta$, $\eta$ and $\lambda$, $\sigma$ jets crowd. The only pronounced exclusion to this pattern seems to be the $\beta$ jet; this case could be, however, assigned to an inadequate resolution of jets in the optical images obtained.

The strong outburst in the detection of fine particles observed to occur between 620 and 720 s after the closest approach at a distance of 4300 km from the nucleus (Tuzzolino et al. 2004) can be attributed (see, e.g., Sekanina et al., 2004) to dissipation of the breaking up chunk split off earlier from the nucleus (leaving aside the question of the reason for and energetics of such a splitting). But one might conceive also of a knot of jets that had intersected earlier. The knot was drifting in space, while the coalescence of fine, 0.01-0.2-µm particles in the ~2 hours of drifting gave rise to the formation of larger, detectable grains with $m \geq 10^{-11}$ g. Interestingly, this outburst in the detection rate exhibits a great excess of particles of just the lowest detectable mass, $m \sim 10^{-11}$ g (see Fig. 4B in Tuzzolino et al., 2004).

## 5. *STARDUST*-returned material

Laboratory examination of the retrieved cometary dust by sophisticated techniques revealed many unexpected findings.

On the whole, the elemental composition of particles agrees fairly well with that of primitive CI meteorites (Flynn et al., 2006). It should be pointed out, however, that practically all data sensitive to temperature and/or oxidation processes suggest involvement of conditions characteristic of combustion under oxygen deficiency.

Let us start with the organics. Sandford et al. (2006) have been reiterating more than once that the polycyclic aromatic hydrocarbons exhibit distributions of two types, with one resembling the products of pyrolysis of the meteoritic macromolecular organics, and the other, high-molecular organic mantles of Interplanetary Dust Particles (IDP; a sizable part of them are assumed to be products of cometary activity), which, in their turn, could again be associated with polymerization under heating during the entry into the Earth's atmosphere (the IDPs are collected on the Earth). The high O and N, and the lower aromatic contents, combined with the enhanced $CH_2/CH_3$ ratios suggest that the *Stardust* organics is in no way similar to the organic material seen in the interstellar diffuse medium (ISM) or in primitive carbonaceous chondrites (Keller et al., 2006). Just as the observed enhanced O/C ratio (Keller et al., 2006), this finding suggests strong difference of Wild 2 organics compared with the ISM product. The organics in the *Stardust* samples is even more "primitive" than that in meteorites, at least in that it is more labile, highly nonuniform, and nonequilibrium.

The isotope composition likewise attests to a high lability of this organics, with the values $\delta^{15}N \approx 1300 \pm 400^0/_{00}$ found by McKeegan et al. (2006) being on a par with the highest values revealed in refractory organics found in IDPs and meteorites.

It appears obvious enough that all the above features can be produced also in oxidizer-deficient combustion of the primitive meteorite organics.

It is often a far from simple task to discriminate between the effects produced in primary heating of components of the Wild 2 nucleus material from the heating caused by the capture (or impact) of high-velocity (6.1 km/s) cometary grains. There is no place for doubt, however, that the dust returned to the Earth contains a certain fraction of crystalline silicates, so that this is not just a set of preserved ISM silicates but rather a mixture of presolar and Solar System materials (Keller et al., 2006; Zolensky et al., 2006). In contrast to the products of impact-induced ejection from Tempel 1 (Lisse et al., 2005, 2006), in the case of Wild 2 no clear-cut evidence for the presence of amorphous silicates, hydrated silicates (phyllosilicates) or carbonates, much less so of ice, has been revealed up to the recent time (Hörz et al., 2006; Flynn et al., 2006; Zolensky et al., 2006), which certainly could appear somewhat strange (Zolensky et al., 2006). And only very scrutinized study by Flynn et al. (2008) has found recently very small grains of the carbonates in Wild 2 dust (as well as their signatures in data obtained for P/Halley in 1986). Viewed from the standpoint considered here, these differences appear only natural. In the case of Wild 2 (and P/Halley), by far the majority of such particles carried out by gas jets, which are high-temperature products of incomplete combustion of $H_2$ + organics, was subjected to a high temperature (~1000 K) decomposing a large share of carbonates and similar compounds, whereas only part of the nucleus material ejected by the impact from Tempel 1 passed through the region of the short-term forced detonation. One could also suggest other reasons for the differences as well. Say, the nuclei of these comets could be ejected in different (e.g., the first or the second) explosions of an envelope, and quite possibly, of different satellites' envelopes even. This suggestion is argued for by the detection of amorphous silicates in the pre-impact coma of Tempel 1 (Lisse et al. 2005). On the other hand, there are indications 'that carbonate also forms by processes other than aqueous alteration in the Solar System' (e.g.,

Flynn et al., 2008 and refs. herein). It is clear that these numerous diverse and subtle effects deserve further studies.

The idea of combustion is corroborated by a somewhat unexpected presence, generally speaking, of particles of two different types, namely, cohesive and dense, on the one hand, and small, fluffy, highly porous particles (Hörz et al., 2006), a finding that could be deduced, as we have pointed out in the preceding Sec. 4, from the analysis of particle distribution in mass performed during the encounter with Wild 2. It is the fluffy particles that would seem earlier to be typical of comets, because they were believed to be primary condensates. It is presently becoming clear, however, that it is the compact particles, probably embedded in an organic mantle produced due to the Fischer-Tropsch catalysis, that are apparently primary, whereas the fluffy particles, the submicron-sized particulate material of soot and "smoke", are actually produced in <u>secondary condensation</u> in the cooling jets formed by the products of incomplete combustion of dirty cometary ice saturated by $2H_2 + O_2$.

## 6. On the origin of Ca-Al-rich inclusions (CAI)

Particular interest is stirred by the discovery in one of the dust grains captured close to Wild 2 of a Ca-Al-rich inclusion (Zolensky et al., 2006). The CAIs occur in carbonaceous chondrites and are believed to be primitive products of the primary high-temperature condensation occurring at $T \approx 1400\text{-}2000$ K (see McKeegan et al., 2006, Zolensky et al., 2006, and refs. therein).

The detection of this inclusion prompted speculations of the need to include large-scale mixing of material in the preplanetary disk which would be capable of transporting particles of the $T \sim 2000$ K condensate from the zone close to the Sun ($R < 0.3$ AU) to the region where comets are assumed to form ($R > 30$ AU) (A'Hearn, 2006). Incidentally, such attempts at explaining the origin of the CAIs and of the originally strongly heated chondrules were undertaken before (see, e.g., Shu et al., 2001; Cuzzi et al., 2003). But in that particular case one considered their transport to the asteroid belt ($R \leq 3$ AU). The requirement of mixing and turbulization of the preplanetary disk *as a whole* brings about a dramatic shortening of its lifetime, a point that calls into question the probability itself of the formation of the giant planets in it (see the old paper by Cameron, 1973); in other words, it practically casts doubt on the disk paradigm as a whole.

One could certainly counter that detection of one such particle does not necessarily entail any implications; indeed, its capture could have no connection with the comet whatsoever and could be caused by a random collision in the dust collection region with a grain ejected in disruption of a suitable meteoroid. A more scrupulous IR spectral analysis of the dust of the Tempel 1 comet performed by Lisse et al. (2006) reveals, however, that the observed fraction of crystallization of the olivine components would require a temperature from 1100 to 1400 K. This temperature is noticeably higher than needed for crystallization of the pyroxenes, $T \approx 800\text{-}900$ K, which could be reached in forced detonation of the electrolyzed ices of the comet (Drobyshevski et al., 2005, 2007). Therefore, the problem of the presence of high-temperature condensates in cometary ices remains unsolved.

It finds, however, a natural solution within the CBC, which considers the Solar system as a limiting case of a binary star. As pointed out in Sec. 2, the CBC turned out capable of (*i*) proposing a solution to a similar problem of the high-temperature origin of the Moon (Drobyshevski, 1975), (*ii*) predicting the existence of a not very remote (~50-300 AU) *joint planeto-cometary cloud* (containing Pluto- or Sedna-type dwarf planets), the source of LP comets, as well as (*iii*) solving a number of other problems as well (for instance, of the origin of the Galilean satellites and of exo-Jupiters on eccentric orbits, etc.).

By this cosmogony, planets (with a possible exclusion of Saturn) formed within a rapidly rotating convective PJ which overflowed its Roche lobe and, therefore, transferred its mass onto the proto-Sun. Starting with the ratio $q = M_J/M_\odot \approx 0.1$, the PJ reached the stage favoring condensation first ($T \approx 2000$ K) of refractory "rocks" and, eventually (at $q < 0.003$-$0.001$), of "ices" (Drobyshevski, 1975, 1978). Because of the turbulization and motion over gravity-dominated trajectories inside the PJ, the products of condensation and their agglomerates (including planetesimals, etc.) could pass through thermodynamically disparate zones to become acted upon over and over again by heat and cold, coalescence and collisional breakup, etc. In the latest stages, at $q \rightarrow 0.001$, when the central temperature in the PJ fell off to $T \approx 300$ K (water condensation), and the PJ started to sink under the surface of its Roche lobe, so that the mass loss rate dropped to $\tau = M_J/(dM_J/dt) \sim 10^4$ years (Drobyshevski, 1978), the conditions became favorable for the formation of MLBs with a high ice content. Four of them became Galilean satellites. We see now that their ices could entrap and preserve minerals and organics disparate in composition and sizes, including the CAI- and ISM-like grains. Ice fragments (i.e., the nuclei of SP comets) created in global explosions of the envelopes of bodies of the type of Galilean satellites should naturally contain, besides the products of electrolysis, particles of this "sand" as well.

## 7. Conclusions

We see that the hopes entertained by the initiators of the *Deep Impact* and *Stardust* projects, namely, that active investigation of comets would shed light on the origin and early stages in the evolution of the Solar System are largely coming true. As always, however, Nature springs surprises on us and prompts researchers to exercise a certain daring and be ready to refine a heretofore universally accepted concept (otherwise why would one need new experiments?).

Our suggestions of a desirable deep survey of the sky with the purpose of discovering numerous trans-Neptune moon- or Pluto-like planets (called presently dwarf planets) moving along disordered orbits, which had been voiced quite a long time ago, as well as of an *in situ* study of SP comets to test the concept of their planetary origin (Drobyshevski, 1978, 1988; Drobyshevski et al., 1995) are gradually seen to materialize. The results obtained in these studies argue convincingly for the validity of the close-binary Solar System cosmogony (based on the statistics of close binaries and present-day understanding of their evolution) and of the paradigm of the planetary origin of comets, an explosive process for SP comets and an impact scenario (in a *joint planeto-cometary cloud*) for the LP comets; a paradigm that actually follows from this cosmogony and has been long expected to come by Lagrange, Vsekhsvyatski and many others who had been taking into account a great diversity of factual material. These new (or very old?) approaches naturally need a comprehensive and thorough examination.

There is presently no doubt whatsoever that the ROSETTA mission to 67P/Churyumov-Gerasimenko will also yield information which, while not fitting the traditional concepts, will find a plausible explanation in terms of the NEC. Particularly, one can predict that the direct measurements will find a high temperature (up to ~900 K) in the jet (sub)surface sources, discover the molecular hydrogen and oxygen dissolved in the cold cometary ices, and give for the nucleus a density of a monolith dirty ice, i.e. about 1000 kg/m$^3$. Missions to the nuclei of new, typical LP comets to reveal their differences from the SP comets would be intriguing but not easy to perform. Indeed, new LP comets could hardly be expected to exhibit, for instance, the distinct jet pattern of outflows, particularly from the non-illuminated part of the nucleus.

We note in conclusion that the already available data amassed in cometary experiments provide serious additional arguments for assigning the highest priority to carrying out comprehensive *in situ* studies of Callisto, which by Drobyshevski et al. (1995) should be aimed at learning the extent of saturation of its massive icy envelope by the products of the volumetric electrolysis and of the associated threat for the Earth.

**Acknowledgements**

The author is grateful to G.J. Flynn of the State University of New York at Plattsburg for valuable discussion and comments concerning properties and composition of cometary nuclei.